\documentclass[onecolumn,useAMS,usenatbib]{mn2e}

\usepackage{amsmath}  
\usepackage{amsfonts} 
\usepackage{graphicx} 
\usepackage{hyperref}
\usepackage{ amssymb }
\usepackage{nicefrac}

\title{A simple exact series representation for relativistic perihelion advance}

\author[Walters]
  {S.J. Walters \\
  School of Mathematics and Physics, University of Tasmania, P.O. Box 37, Hobart, 7001, Tasmania, Australia}

\date{\today}

\begin{document}

\maketitle 

\begin{abstract}
The small discrepancy between the observed orbit of Mercury and the orbit predicted by Newtonian gravity was a key test of Einstein's theory, and a dramatic verification of the correctness of General Relativity. This `anomalous precession' of the perihelion is here calculated using a particularly simple method and resulting in a new and elegant series for the precession which converges very quickly.
\end{abstract}

\begin{keywords}
gravitation, celestial mechanics.
\end{keywords}

\section{Introduction}

The anomalous precession of Mercury's perihelion of approximately $43$ arc-seconds per century is one of the classic tests of the General Theory of Relativity, and a very significant support for the theory. This short note lays out a simple, exact series representation for calculating the value of this precession for a small mass orbiting a spherically symmetric body. We are here concerned with the relativistic effects only, and not the precession due to other orbiting masses \citep{Stewart}, solar oblateness \citep{xu} or any other effects \citep{Lo}.

\section{Analytic calculation}
\label{part3}
In order to evaluate the angular change for a satellite going from perihelion to aphelion we begin with the Schwarzschild metric for the equatorial plane. Using the (-,+,+,+) convention, the corresponding line element equation is
\begin{eqnarray}
ds^2=-c^2 d\tau^2=-c^2 dt^2(1-r_s/r)+\frac{dr^2}{1-r_s/r}+r^2 d\phi^2\nonumber
\end{eqnarray}
where the central mass has Schwarzschild radius $r_s=2MG/c^2$.
Parameterizing by proper time $\tau$, and indicating derivatives with respect to $\tau$ using dot notation, this may be written 
\begin{eqnarray}
\dot s^2=-c^2=-c^2 \dot t^2(1-r_s/r)+\frac{\dot r^2}{1-r_s/r}+r^2 \dot\phi^2.
\label{eqn1}
\end{eqnarray}

The equations of motion are derived by solving the Euler-Lagrange equations for extremizing proper distance, $s$. The Lagrangian has no direct dependence on $t$ or $\phi$ so the first integrals for each of these can be written immediately, giving the following two equations:
\begin{eqnarray}
\dot t&=&\frac{E}{1-rs/r}\nonumber\\
\dot \phi&=&\frac{c L}{r^2}
\label{eqn2}
\end{eqnarray}
for constants of integration $E$ and $L$. Rather than solve the Euler-Lagrange equation for $r$, it is easier to substitute these equations for $\dot t$ and $\dot \phi$ into equation (\ref{eqn1}) and solve for $\dot r$. Dividing $\dot r$ by $\dot \phi$ (equation (\ref{eqn2})), gives a first order differential equation for the path of the planet:
\begin{eqnarray}
\frac{dr}{d\phi}=\pm r^2\sqrt{\frac{E^2-1}{L^2}+\frac{r_s}{L^2 r}-\frac{1}{r^2}(1-\frac{r_s}{r})}.\nonumber
\end{eqnarray}
The constants $E$ and $L$ are related to the energy and angular momentum of the satellite. However, for ease of calculations here, it is helpful to define new constants to bring the form of the equation to 
\begin{eqnarray}
\frac{dr}{d\phi}=\pm r^2\sqrt{C_1+C_2 \frac{1}{r}-\frac{1}{r^2}(1-\frac{r_s}{r})},
\label{eqn4}
\end{eqnarray}
in which the new constants are $C_1=\frac{E^2-1}{L^2}$; $C_2=\frac{r_s}{L^2}$. Note that this differs from the Newtonian path equation $\frac{dr}{d\phi}=\pm r^2\sqrt{C_1+C_2 \frac{1}{r}-\frac{1}{r^2}}$ in the last term only. Next, these constants $C_1$ and $C_2$ may be determined in terms of known parameters. These are the point of closest approach to the sun (perihelion), denoted $r_p$, and the point furthest from the sun (aphelion), $r_a$. At these points, the radius $r$ is a minimum or maximum, so $\frac{dr}{d\phi}=0$. The equation for $r=r_p$ is then:
\begin{eqnarray}
\frac{dr}{d\phi}\vline_{(r=r_p)} =0=\pm r_p^2\sqrt{C_1+C_2 \frac{1}{r_p}-\frac{1}{r_p^2}(1-\frac{r_s}{r_p})},\nonumber
\end{eqnarray}
which may be solved for $C_1$. Replacing $C_1$ in equation (\ref{eqn4}) with this solution, we may then take this new equation and put $r=r_a$, which is the other turning point, so $\frac{dr}{d\phi}=0$ there also. Solving this for $C_2$ gives the path equation for the planet, fully specified in terms of the mass of the sun (as $r_s=2MG/c^2$), and the distances from the sun at perihelion and aphelion:
\begin{eqnarray}
\frac{dr}{d\phi}=\pm r^2\sqrt{\frac{1}{r_p}-\frac{1}{r}}\sqrt{\frac{1}{r}-\frac{1}{r_a}}\sqrt{1-r_s(\frac{1}{r}+\frac{1}{r_a}+\frac{1}{r_p})}.\nonumber
\end{eqnarray}
The `$\pm$' sign relates only to the direction of travel about the sun, which is unimportant for this calculation, being concerned only with the magnitude of the perihelion shift. As the planet moves from perihelion $r_p$, to aphelion $r_a$, the change in $\phi$ may be written as
\begin{eqnarray}
\triangle\phi=\int {d\phi}=\int_{r_p}^{r_a}{\frac{d\phi}{dr}dr}=\int_{r_p}^{r_a}{\frac{1}{dr/d\phi}dr}=\int_{r_p}^{r_a}{\frac{dr}{r^2\sqrt{\frac{1}{r_p}-\frac{1}{r}}\sqrt{\frac{1}{r}-\frac{1}{r_a}}\sqrt{1-r_s(\frac{1}{r}+\frac{1}{r_a}+\frac{1}{r_p})}}}\nonumber
\end{eqnarray}

This integrand is clearly singular at $r=r_a$ and $r=r_p$, but this may be resolved by two simple substitutions. First we let $r=\nicefrac{1}{u}$ and introduce the constants $u_a=\nicefrac{1}{r_a}$ and $u_p=\nicefrac{1}{r_p}$. In the resulting integrand, completing the square suggests the substitution $(u-\frac{u_a+u_p}{2})=(\frac{u_p-u_a}{2})\sin \theta$. Making the substitution and changing the limits accordingly we obtain
\begin{eqnarray}
\triangle\phi&=&\int_{-\pi/2}^{\pi/2}{\frac{d \theta}{\sqrt{1-r_s[\frac{u_p-u_a}{2}\sin \theta+3\frac{u_a+u_p}{2}]}}}.\nonumber
\end{eqnarray}
At this point, the integrand is made more readable by introducing the standard orbital parameters of eccentricity $e$, semi-major axis $a$, and semilatus rectum $p$, related to radii at perihelion and aphelion by $r_p=a(1-e)$, $r_a=a(1+e)$ and $p=a(1-e^2)$. Incidentally, we note that using these identities, we can work backwards to identify the original constants $E$ and $L$ by
\begin{eqnarray}
E^2&=&1-\frac{r_s}{2a}\frac{1-2r_s/p}{1-r_s(3+e^2)/(2p)}\nonumber\\
L^2&=&\frac{r_s p/2}{1-r_s(3+e^2)/(2p)},\nonumber
\end{eqnarray}
fully specifying the equations of motion in terms of the orbital parameters, as found in standard texts (for example, \citet{pois}, pp.272-274).

In terms of these new constants, $\nicefrac{1}{2} (u_a+u_p)=\nicefrac{1}{p}$ and $\nicefrac{1}{2} (u_p-u_a)=\nicefrac{e}{p}$ so the integral becomes
\begin{eqnarray}
\triangle\phi&=&\int_{-\pi/2}^{\pi/2}{\frac{d \theta}{\sqrt{1-\frac{3r_s}{p}-\frac{r_s e}{p}\sin \theta}}}\nonumber\\
&=&\frac{1}{\sqrt{1-3r_s/p}}\int_{-\pi/2}^{\pi/2}{\frac{d \theta}{\sqrt{1-\frac{r_s e}{p}\frac{\sin \theta}{1-3 r_s/p}}}},\nonumber
\end{eqnarray}
where the factor in front of the integral is the precession in the small eccentricity limit (see \citet{pois}, p.275). We now have an integral which cannot be evaluated in terms of elementary functions (although it is a complete elliptic integral, \cite{abram}, p.589). However, this integral can be evaluated to arbitrary precision, by using for example, Gaussian quadrature. Alternatively, we can obtain a power series representation using the binomial expansion $(1-x)^{-1/2}=1+\frac{1}{2}x+\frac{3}{8}x^2+\frac{5}{16}x^3+\frac{35}{128}x^4+...$ Defining a new constant $\beta=\frac{r_s e}{p-3 r_s}$, this integral may be written as the series
\begin{eqnarray}
\triangle\phi&=&\frac{1}{\sqrt{1-3r_s/p}}\int_{-\pi/2}^{\pi/2}{\bigg[1+\frac{1}{2}\beta \sin \theta+\frac{3}{8}\beta^2 \sin^2 \theta+... \bigg]d\theta}.\nonumber
\end{eqnarray}
Each term in this series may be integrated analytically, with all the odd powers of $\sin\theta$ integrating to zero over the interval. After evaluating the integrals of $\sin^{2n}(\theta)$, the total change in azimuthal angle $\phi$ in going from perihelion to aphelion is thus given by
\begin{eqnarray}
\triangle\phi&=&\frac{\pi}{\sqrt{1-3r_s/p}} \bigg[1+\sum_{n=1}^{\infty}\frac{(4n)!}{n!^2(2n)! 2^{6n}} \beta^{2n}\bigg].
\label{eqn7}
\end{eqnarray}

The Newtonian value and Einstein's approximate correction are given by the expansion of equation (\ref{eqn7}) in powers of $r_s/p$ to give $\triangle\phi_0=\pi$ and $\triangle\phi_{1}=\frac{3 r_s \pi}{2 p}$ as given, for example, by \cite{chandra}, p.108. Note that this is for a half revolution, so multiplying by two gives the precession in radians per revolution. The approximate Schwarzschild radius of the sun is $r_s=\frac{2 G M}{c^2}=2.95$ km. For Mercury, $r_p=46,001,200$ km, $r_a=69,816,900$ km, and there are approximately $415.2$ revolutions per century. Converting from radians to arc-seconds gives the result $\approx 43$ arc-seconds per century.

After subtracting the Newtonian value of $\pi$, the standard first order approximation $\frac{3 r_s \pi}{2 p}$ radians per half revolution gives the correct figure to seven significant figures, while the first term of equation (\ref{eqn7}),$\frac{\pi}{\sqrt{1-3r_s/p}}-\pi$ gives nine correct significant figures. Using just the first two terms
\begin{eqnarray}
\triangle\phi-\pi&\approx&\frac{\pi}{\sqrt{1-3r_s/p}}(1+\frac{3}{16}\beta^2)-\pi
\label{eqn8}
\end{eqnarray}
gives 25 correct significant figures! The size of the constant $\beta$ is primarily a product of two dimensionless numbers: the eccentricity $e$ and the ratio $r_s/p$. In the case of Mercury's orbit, $e\approx0.2$ and $r_s/p\approx5\times10^{-8}$ so $\beta^2\approx10^{-16}$. Thus, each additional term adds approximately sixteen decimal places of accuracy.

\section{Conclusion}
The calculation of the anomalous precession of Mercury has been presented by means of a straight-forward derivation of the path equation in the Schwarzschild space-time utilizing the Euler-Lagrange equations. The result of this calculation is an integral that can be solved numerically to arbitrary precision. Alternatively, by expanding the integrand, the result can be expressed analytically as a simple, elegant infinite series. In the case of Mercury's orbit, this series converges to the correct value at a rate of approximately sixteen decimal places per term. Thus the simple two-term approximation (\ref{eqn8}) for $\triangle\phi$ gives an accessible and highly accurate formula for use by practitioners.

\section*{Acknowledgements}
I am indebted to Professor Larry Forbes and Professor Clifford Will for their helpful comments on earlier drafts of this paper.

\end{document}